\documentclass[twocolumn,prl,superscriptaddress,showpacs,floats]{revtex4}
\usepackage{graphicx}
\usepackage{amsmath}

\newcommand{\Mnsub}{Mn$_{\mathrm{sub}}$}
\newcommand{\Mnint}{Mn$_{\mathrm{int}}$}
\begin{document}
\title{Density-functional theory  study of half-metallic heterostructures: interstitial Mn in Si}
\author{Hua Wu}
\affiliation{Fritz-Haber-Institut der Max-Planck-Gesellschaft,
Faradayweg 4-6, D-14195 Berlin, Germany}
\affiliation{II. Physikalisches Institut, Universit\"{a}t zu
K\"{o}ln, Z\"{u}lpicher Str. 77, D-50937 K\"{o}ln, Germany}
\author{Peter Kratzer}
\affiliation{Fachbereich Physik, Universit{\"a}t Duisburg-Essen,
Lotharstr. 1, D-47048 Duisburg, Germany}
\author{Matthias Scheffler}
\affiliation{Fritz-Haber-Institut der Max-Planck-Gesellschaft,
Faradayweg 4-6, D-14195 Berlin, Germany}
\begin{abstract}
Using density-functional theory within the generalized gradient
approximation, we show that Si-based heterostructures with 1/4
layer $\delta$-doping of {\em interstitial} Mn (Mn$_{\mathrm
int}$) are half-metallic. For Mn$_{\mathrm int}$ concentrations of
1/2 or 1 layer, the states induced in the band gap of
$\delta$-doped heterostructures still display high spin
polarization, about 85\% and 60\%, respectively. The proposed
heterostructures are more stable than previously assumed
$\delta$-layers of {\em substitutional} Mn. Contrary to
wide-spread belief, the present study demonstrates that {\em
interstitial} Mn can be utilized to tune the magnetic properties
of Si, and thus provides a new clue for Si-based spintronics
materials.
\end{abstract}
\maketitle
%
Doping of semiconductors with magnetic $3d$ transition metal
impurities has attracted much interest: recent advances in
spintronics have been largely due to magnetic semiconductors
produced in this way.
Their spin-dependent electric transport properties ~\cite{Zutic04}
could be exploited in novel magneto-electronic devices, e.g. spin
filters.
Since it has long been known that manganese impurities can have a
high magnetic moment~\cite{LW62,BAS85}, research has focused
Mn-doped magnetic semiconductors based on GaAs and
Ge~\cite{Ohno98,Dietl00,Park02}.
Moreover, doping of GaAs by a $\delta$-layer of Mn has been
proposed theoretically~\cite{Sanvito01}, and an experimental
realization has been reported recently~\cite{Nazmul05}.
The calculations of spin-polarized transport\cite{Sanvito01}
employing density-functional theory (DFT) within the
local-spin-density approximation (LSDA) and pseudopotentials
predicted that GaAs doped with a $\delta$-layer of substitutional
Mn (\Mnsub) on Ga sites is a two-dimensional (2D) ferromagnetic
half-metal with large exchange coupling. Indeed, a Curie
temperature of up to 250~K has been measured on Mn $\delta$-doped
GaAs samples~\cite{Nazmul05}.
Adding magnetic functionality to the most common semiconductor,
Si, is still in its infancy, despite recent reports about
ferromagnetism in Mn-doped Si created by ion
implantation~\cite{Bolduc05}.  Recently, Qian {\it et
al.}~\cite{Qian06} proposed, on the basis of pseudopotential
calculations within the generalized gradient approximation (GGA),
that a Si heterostructure with \Mnsub~ $\delta$-doping should be a
2D ferromagnetic half-metal.
So far, research on Mn-doped Si has concentrated on {\em
substitutional} Mn~\cite{Stroppa03,Picozzi04}, probably motivated
by the physics of dilute magnetic semiconductors, where it has
been argued that \Mnsub~ is crucial for
ferromagnetism~\cite{Dietl00,Park02,Dietl01,Erwin02}. However,
these substitutional Mn impurities in Si are energetically less
stable than interstitial ones~\cite{Dalpian03,Luo04,Silva04}.
This has been considered a serious obstacle in creating magnetic
Mn:Si calling for growth techniques far from
equilibrium~\cite{Dalpian03}.

In this Letter, we investigate the role of {\em interstitial} Mn
(\Mnint) impurities for ferromagnetism in Si by means of
all-electron full-potential DFT calculations. We propose a novel
type of heterostructures with \Mnint~ $\delta$-doping. For
occupation of each forth interstitial site in a single layer by
Mn, we find 2D half-metallic behavior with a Kohn-Sham gap of
0.5~eV in the electronic density of states in the {\em majority}
spin channel. Even for higher doping densities, the
spin-polarization is still large, $>$60\%, judged from the density
of states at the Fermi level.
This is different from the previously proposed \Mnsub~
$\delta$-layers~\cite{Qian06}, where a (smaller) gap in the {\em
minority} spin channel between Mn states of $e$ and $t_2$-type
symmetry was responsible for the half-metallicity.
Moreover, our proposed \Mnint~ $\delta$-layer is calculated to be
$~0.5$~eV per Mn atom lower in energy than the heterostructure
proposed by Qian {\it et al.}~\cite{Qian06}, in line with the
known higher stability of isolated \Mnint~ impurities compared to
\Mnsub~ impurities in Si~\cite{Luo04,Silva04}.
Consequently, Si self-interstitials can destroy the
half-metallicity of the heterostructure assumed by Qian {\it et
al.} by energetically favorable site exchange with \Mnsub.

\begin{figure}[t]
\centering \includegraphics[width=8.3cm]{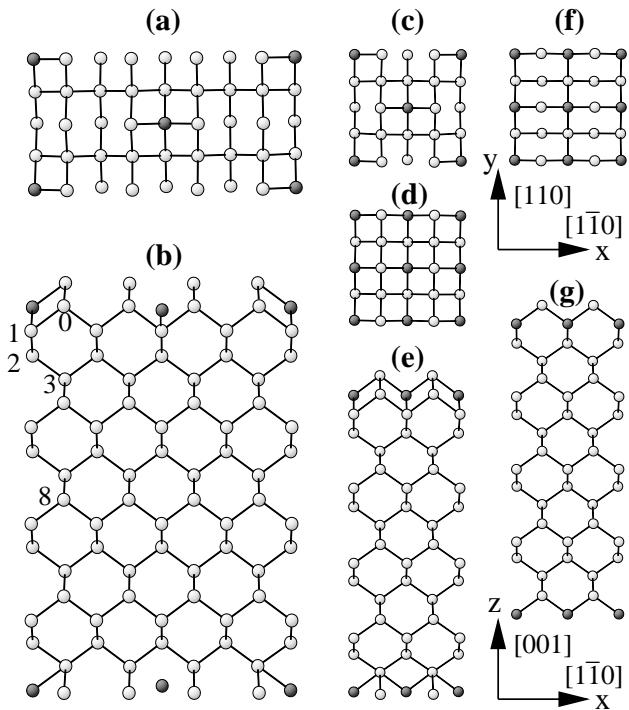} \caption{(a)
(001)-$xy$-plane view and (b) (110)-$xz$-plane view of the
Si-based heterostructure (modeled by a 16-layer supercell) with
1/4 layer of interstitial Mn in  $c(4\times 2)$ arrangement; (c)
(001)-plane view of the heterostructure with 1/2 layer of
interstitial Mn in $p(2\times 2)$ arrangement; (d) (001)-plane
view and (e) (110)-plane view of the heterostructure with a full
layer of interstitial Mn in $(1\times 1)$ arrangement; and (f)
(001)-plane view and (g) (110)-plane view of the heterostructure
with a full layer of substitutional Mn in $(1\times 1)$
arrangement. Black balls show Mn and gray balls Si atoms. The
silicon atoms in different layers are marked with indices.}
\end{figure}

Our proposed Si-based heterostructures are modeled by a Si(001)
supercell with 16 atomic layers (see Fig.~1). Such thickness
ensures that the central Si layer behaves as in the bulk, and the
interaction between Mn layers is negligible. Mn deposited on a
Si(001) surface occupies preferably subsurface interstitial
sites~\cite{Wu04}, and a $c(4\times 2)$ arrangement for 1/4
monolayer (ML) \Mnint~ has been identified recently by a combined
scanning tunneling microscopy (STM) and DFT
study~\cite{LaBella06}. Therefore we first study the $c(4\times
2)$ heterostructure consisting of 16-layer Si and 1/4~ML \Mnint~
in $c(4\times 2)$ arrangement, as seen in Figs.~1(a) and 1(b). We
also perform calculations for 1/4~ML \Mnint~ in $p(2\times 2)$
arrangement (not shown). Secondly, we deal with a $p(2\times 2)$
heterostructure with 1/2~ML \Mnint, as shown in Fig.~1(c). Finally
we compare the $(1\times 1)$ heterostructure having 1 ~ML \Mnint~
(Figs.~1(d) and 1(e)) with the $(1\times 1)$ heterostructure
having 1~ML \Mnsub~ (Figs.~1(f) and 1(g)) studied by Qian {\it et
al.}~\cite{Qian06}.

We perform DFT-GGA \cite{Perdew} calculations using the
full-potential augmented plane-wave plus local-orbital
method~\cite{Blaha}, which was shown to be appropriate for
description of the Mn:Si system~\cite{Wu04}. The lattice constant
of 5.48~{\AA} for bulk Si calculated in GGA is used to construct
the heterostructures we study. The muffin-tin sphere radii are
chosen to be 1.11~{\AA} for both the Mn and Si atoms. Converged
results are obtained at a cut-off energy of 13.8~Ryd for the
interstitial plane-wave expansion, and with a set of
12$\times$12$\times$1 special $\bf{k}$-points for integrations
over the Brillouin zone of the $(1\times 1)$ heterostructure (and
an equivalent density of $\bf{k}$-points for larger supercells,
e.g., 6$\times$5$\times$1 for the $c(4\times 2)$ cell). All Si and
Mn atoms except for the central Si layer are relaxed until the
calculated atomic force for each of them is smaller than
0.05~eV/{\AA}.

The $c(4\times 2)$ heterostructure (see Figs.~1(a) and 1(b))
displays a half-metallic electronic band structure as shown by the
density of states (DOS) derived from the Kohn-Sham energy levels
(see Fig.~2(a)). The spin-up channel (dashed blue lines) is
insulating with a Kohn-Sham gap of 0.5~eV. The spin-down channel
(solid red lines) is metallic, and the Fermi level lies 0.2~eV
below the bottom of the spin-up conduction bands. While the Mn
$3d$ spin-up orbitals are fully occupied, the planar spin-down
$x^2-y^2$ orbitals form a conduction band with bandwidth of
0.7~eV, contributing to the DOS in the metallic channel. The
narrower $xz$- and $yz$-derived conduction bands are also
partially occupied and add to the DOS at the Fermi
energy~\cite{footnote}. The coplanar Si (Si$_0$), the
first-neighbor-layer Si (Si$_1$) and the second-neighbor-layer Si
(Si$_2$) (cf. Fig.~1(b)) also contribute considerably to the
half-metallic DOS through Mn $3d$-Si $3s3p$ hybridizations. Such
contribution decreases strongly for the third-neighbor-layer Si
(Si$_3$) and eventually vanishes for the central-layer Si
(Si$_8$). The Mn atom has a local spin magnetic moment (within the
muffin-tin sphere) of 2.56~$\mu_B$, and its four Si neighbors
Si$_1$ (to which it is tetrahedrally coordinated) get
spin-polarized with a spin moment of about 0.024~$\mu_B$ each. Any
other Si atom has a spin moment less than 0.01 $\mu_B$. Taking
also into account the spin moment of 0.25~$\mu_B$ outside the
muffin-tin spheres, stemming from the Mn $3d$-Si $3s3p$ hybridized
bands, the total spin moment is an integer of 3 $\mu_B$ per
supercell \cite{Note}. Note that due to the long distance of
8.7~{\AA} between the center and corner Mn atom(s) in Fig.~1(a)
the magnetic coupling turns out to be weak, but
ferromagnetic~(FM): Our calculations show that the FM state is
slightly more stable than the antiferromagnetic (AF) one by 5
meV/Mn, consistent with the short range of FM interactions in
zincblende MnSi and MnGe~\cite{sasioglu:05,picozzi:06}.

\begin{figure}[t]
\centering \includegraphics[width=9.0cm]{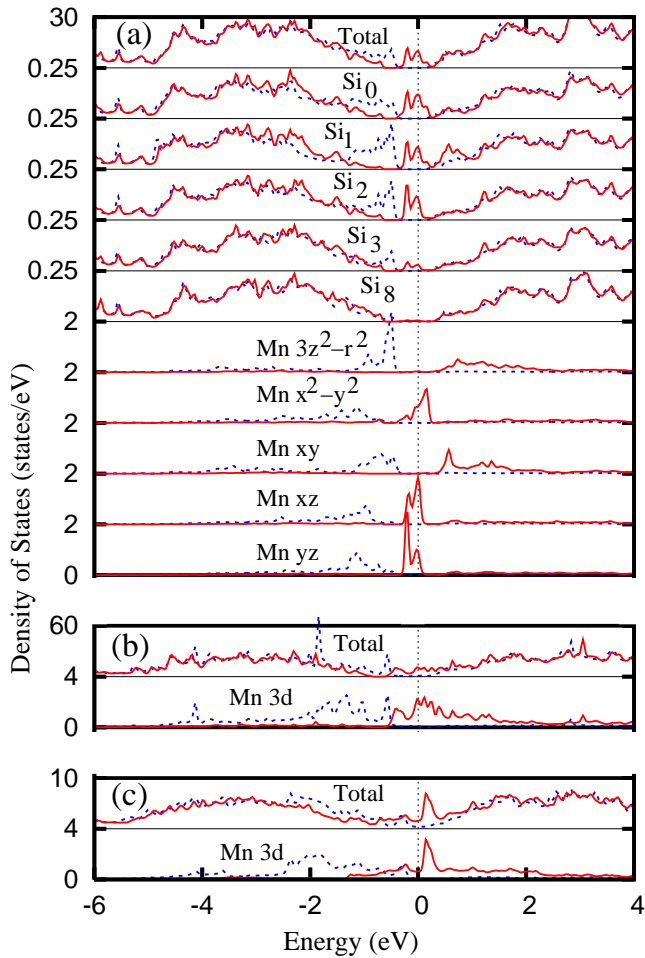} \caption{
(Color online) DOS of (a) the 1/4-layer interstitial Mn $c(4\times
2)$, (b) the 1/2-layer Mn $p(2\times 2)$, and (c) the 1-layer Mn
$(1\times 1)$ heterostructures [cf. Figs. 1(a)-1(e)]. The solid
red (dashed blue) line refers to the down (up) spin component. The
Fermi level is set to zero. (a) shows half-metallicity, (b) and
(c) a high spin-polarization at the Fermi level.}
\end{figure}

The 1/4 ML $p(2\times 2)$ heterostructure (not shown, but refer to
Fig. 1(c) and remove the central Mn in the $p(2\times 2)$ cell)
also displays half-metallicity, similar to  the above discussed
1/4 ML $c(4\times 2)$ heterostructure. The spin-down $x^2-y^2$
orbitals hybridized with Si $sp$ and the spin-down $xz$ and $yz$
orbitals determine the half-metallic DOS (not shown). The
$p(2\times 2)$ superstructure is only sightly less stable than the
above $c(4\times 2)$ one by 20 meV per supercell, which shows that
the $c(4\times 2)$ superstructure gains a little more by lattice
relaxation.

For the 1/2 ML $p(2\times 2)$ heterostructure (see Fig. 1(c)), the
DOS is displayed in Fig. 2(b), indicating a high spin-polarization
at the Fermi level of $~85$\%. The Mn spin moment is 2.72~$\mu_B$
and the total one is 6.34~$\mu_B$ per supercell (3.17~$\mu_B$ per
Mn). Compared with the $c(4\times 2)$ supercell, the conduction
bands become broader due to decreased Mn-Mn distance and enhanced
Mn $d$-Si $sp$ hybridization. The FM exchange coupling of Mn to
its four first neighbors at 5.5~{\AA} is quite large, $J=13$~meV.
This estimate is based on a Heisenberg model with local Mn spin of
$S=3/2$, and the calculated stability of the FM state over the AF
state $4JS^2=120$~meV per Mn atom. Thus the 1/2~ML $p(2\times 2)$
heterostructure combines the advantage of a stronger in-plane FM
coupling with a still high spin-polarization.

\begin{figure}[t]
\centering \includegraphics[width=8.0cm]{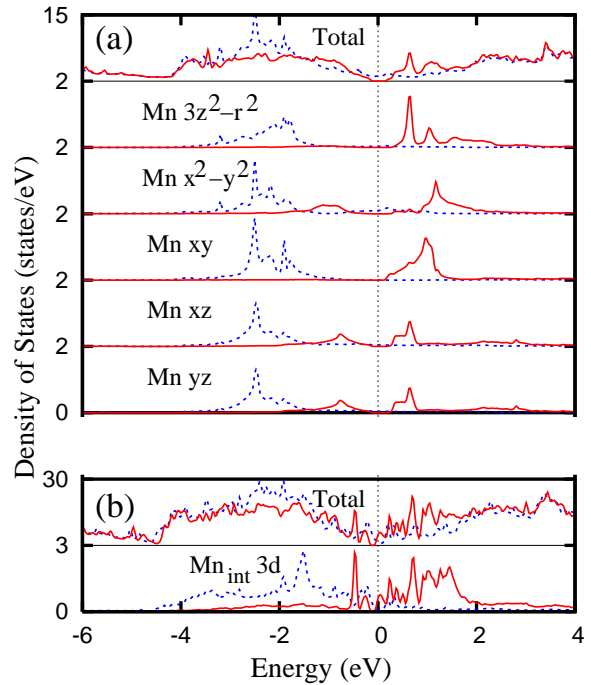} \caption{
(Color online) DOS of the 1-layer substitutional Mn. (a):
$(1\times 1)$ heterostructure shows half-metallicity; (b):
$p(2\times 2)$ heterostructure with $extra$ 1/4-layer co-planar
interstitial Si shows, after an energetically favorable Si-Mn
site-interchange, the loss of half-metallicity. The solid red
(dashed blue) line refers to the down (up) spin component.}
\end{figure}

Now we turn to the 1~ML $(1\times 1)$ heterostructure (see Figs.
1(d) and 1(e)). Even in this case, spin-polarization is still
$>60$\%, as shown by our calculations (Fig. 2(c)). The Mn and Si
atoms form broad conduction bands. The Mn spin moment is
2.68~$\mu_B$ and the total one is 3.12~$\mu_B$ per $(1\times 1)$
supercell. According to our calculations, the Mn-Mn in-plane FM
coupling turns out to be further increased: the FM state is more
stable than the AF state by 570~meV/Mn. Since the direct magnetic
coupling between two neighboring Mn atoms (with distance of
3.9~{\AA}) is AF, we attribute the increased FM coupling with
shorter Mn--Mn distance to an enhanced interaction of
double-exchange type. It operates in the impurity band formed by
the spin-down $3d$ states of Mn with $t_2$ character, as seen in
Fig. 2, and is mediated by spin-down Si $sp$ states. As a result,
we find the Si atoms to have small spin moments parallel to Mn in
all the above calculations.

Next we show that the heterostructure with \Mnsub~ (Figs.~1(f) and
1(g)) proposed by Qian {\it et al.}~\cite{Qian06} is energetically
unfavorable. Fig.~3(a) shows that our calculations yield
half-metallic behavior with a Kohn-Sham gap of 0.25~eV for 1~ML
\Mnsub~ $(1\times 1)$, in agreement with Qian {\it et
al.}~\cite{Qian06}. The Mn spin moment is 2.97~$\mu_B$, each
first-neighbor Si has a spin moment of $-0.116 \mu_B$, and the
total one is again integer, 3~$\mu_B$ per $(1\times 1)$ supercell.
However, the \Mnsub~ heterostructure turns out to be less stable
than the \Mnint~ one by 490~meV per $(1\times 1)$ supercell,
assuming bulk Si as reservoir for the extra Si.
Knowledge about Mn-Si compounds corroborates the metastability of
tetrahedrally coordinated \Mnsub: Mn is known to prefer high
coordination to Si, as exemplified by the seven-fold Mn-Si
coordination in the ground-state crystal structure of MnSi, or the
eight-fold coordination in the CsCl structure~\cite{Wu04}. In
contrast, the hypothetic zinc-blende structure of MnSi with
tetrahedral coordination of Mn to Si is calculated to be extremely
unstable, 2.3~eV higher in energy than the ground state.  The
heterostructures with \Mnint~ proposed by us, having four
second-neighbor Si atoms (2.7~{\AA}) in addition to four
first-neighbor Si atoms (2.4~{\AA}), comply with the rules for the
optimal Mn-Si coordination derivable from bulk phase stability.

We note that both types of $\delta$-layers considered are only
metastable in the thermodynamic sense, i.e. they could be
destroyed by annealing. However, the tendency of Mn to form
ordered subsurface layers, as observed recently~\cite{LaBella06},
suggests that growth kinetics could be exploited to fabricate the
\Mnint~ heterostructure proposed by us. Although the reported
Mn-induced $c(4\times 2)$ structure has so far been observed only
in islands, extended layers might be obtainable by improved growth
techniques, and could be overgrown at lower temperatures with a
capping layer.
After successful preparation, the heterostructures are still
jeopardized by degradation due to highly mobile~\cite{Centoni05}
Si self-interstitials: By studying the 1~ML \Mnsub~ $p(2\times 2)$
heterostructure with an {\em extra} 1/4~ML co-planar interstitial
Si, we find that interstitial Si is capable of site exchange with
one \Mnsub, thereby considerably lowering its energy by 1.2~eV.
This irreversibly destroys the half-metallicity of the
\Mnsub-heterostructure (cf. Ref.~\onlinecite{Qian06}) by degrading
the spin-polarization at the Fermi level to about 40\%, as seen in
Fig. 3(b). In the \Mnint-heterostructure, trapped co-planar Si
interstitials have a less drastic effect on the electronic
structure, but they could shift the Fermi level out of the
half-metallic gap if present in high concentrations.

\begin{figure}[t]
\centering \includegraphics[width=7.0cm]{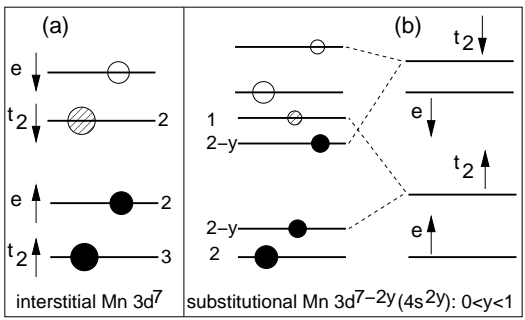} \caption{
Schematic view of the electronic states of the (a) interstitial
and (b) substitutional Mn in Si~[\onlinecite{footnote}]. The
filled, patterned, and open circles indicate full, partial, and
zero occupation, respectively. Both the size of the circles and
the numbers next to the levels indicate their occupation number.
The half-metallic DOS results in (a) from the $t_{2\downarrow}$
band, but in (b) from the $t_{2\uparrow}$ antibonding band.}
\end{figure}

We stress that the half-metallicity in our proposed
\Mnint-heterostructure (metallic minority spin channel) and in the
\Mnsub-heterostructure of Qian {\it et al.} (metallic majority
spin channel) have different physical origin. This is explained in
the level diagram of Fig.~4. The \Mnint~ $3d$ states are
energetically located near the top of the Si-host valence bands
(Fig. 2(a)). The weak crystal field with $T_d$ symmetry and strong
Hund exchange stabilize a high-spin state for
\Mnint~\cite{LW62,BAS85}. The $4s\rightarrow$$3d$ electron
transfer leads to a $3d^7$
($t_{2\uparrow}^{3}e_{\uparrow}^{2}t_{2\downarrow}^{2}$)
configuration (Fig. 4(a)). Thus the partially filled
$t_{2\downarrow}$ band is responsible for the metallic spin-down
DOS seen in Fig.~2(a). For \Mnsub, however, the Mn states of $t_2$
symmetry mix strongly with Si $sp$ states of the same symmetry,
thus splitting the $t_2$ states into a bonding and an antibonding
band, lying in the valence and in the conduction band of the Si
host, respectively (c.f. Fig.~3(a)). The \Mnsub~ $3d$
$t_{2}^{4-2y}$ and $4s^{2y}$, in total four electrons, fill up the
bonding states, together with the Si $sp$ electrons (Fig. 4(b)).
Note that the $4s\rightarrow 3d$ electron transfer in the \Mnsub~
is not as complete as in the \Mnint, as our calculations find the
\Mnsub~ $3d$ occupation (within the muffin-tin sphere) to be less
than the \Mnint~ one by 0.2 electrons.
Since the non-bonding $e_{\uparrow}$ states (lying below the
$t_{2\uparrow}$ states for $T_d$ symmetry) are fully occupied by
two electrons, the remaining one electron resides in the
$t_{2\uparrow}$ antibonding band that gives rise to the DOS in the
spin-up channel in Fig.~3(a).

To conclude, we proposed a stable Si-based heterostructure using
$\delta$-doping by interstitial Mn, and proved it to be
half-metallic for 1/4 monolayer of Mn, using DFT calculations. A
recent combined STM-DFT study~\cite{LaBella06} lets us expect that
such structures could be prepared by molecular beam epitaxy. In
order to achieve ferromagnetic ordering at room temperature, one
should aim at higher Mn concentrations. Even for a full
$\delta$-layer of \Mnint, the spin polarization of the conduction
electrons should still be as high as 60\%.


\begin{thebibliography}{15}
\bibitem{Zutic04} For a review, see e.g.
I.~\v{Z}uti\'{c}, J.~Fabian, and S.~D. Sarma,
\newblock Rev. Mod. Phys. {\bf 76}, 323 (2004).

\bibitem{LW62}
G. W. Ludwig and H. H. Woodbury, Solid State Phys.
{\bf 13}, 223 (1962).

\bibitem{BAS85}
F. Beeler, O. K. Andersen, and M. Scheffler, Phys. Rev. Lett. {\bf
55}, 1498 (1985); Phys. Rev. B {\bf 41}, 1603 (1990).

\bibitem{Ohno98}
H. Ohno, Science {\bf 281}, 951 (1998).

\bibitem{Dietl00}
T. Dietl, H. Ohno, F. Matsukura, J. Cibert, and F. Ferrand,
Science {\bf 287}, 1019 (2000).

\bibitem{Park02}
Y. D. Park, A. T. Hanbicki, S. C. Erwin, C. S. Hellberg, J. M. Sullivan, J. E. Mattson,
T. F. Ambrose, A. Wilson, G. Spanos, and B. T. Jonker,
Science {\bf 295}, 651 (2002).

\bibitem{Sanvito01} S. Sanvito and N.A. Hill, Phys. Rev. Lett. {\bf 87}, 267202 (2001).

\bibitem{Nazmul05} A.M. Nazmul, T. Amemiya, Y. Shuto, S. Sugahara, and M. Tanaka,
Phys. Rev. Lett. {\bf 95}, 017201 (2005); $ibid.$ {\bf 96}, 149901(E) (2006).

\bibitem{Bolduc05} M. Bolduc, C. Awo-Affouda, A. Stollenwerk, M.B. Huang, F.G. Ramos,
G. Agnello, and V.P. LaBella, Phys. Rev. B {\bf 71}, 033302
(2005).

\bibitem{Qian06} M.C. Qian, C.Y. Fong, K. Liu, W.E. Pickett, J.E. Pask, and L.H. Yang,
Phys. Rev. Lett. {\bf 96}, 027211 (2006).

\bibitem{Picozzi04}
S. Picozzi, F. Antoniella, A. Continenza, A. MoscaConte, A.
Debernardi, and M. Peressi,
\newblock Phys. Rev. B {\bf 70}, 165205 (2004).

\bibitem{Stroppa03}
A. Stroppa, S. Picozzi, A. Continenza, and A. J. Freeman,
\newblock Phys. Rev. B {\bf 68}, 155203 (2003).

\bibitem{Dietl01}
T. Dietl, H. Ohno, and F. Matsukura
\newblock Phys. Rev. B. {\bf 63}, 195205 (2001).

\bibitem{Erwin02} S.C. Erwin and A. G. Petukhov,
Phys. Rev. Lett. {\bf 89}, 227201 (2002).

\bibitem{Dalpian03}
G.~M. Dalpian, A.~J.~R. da Silva, and A. Fazzio,
\newblock Phys. Rev. B {\bf 68}, 113310 (2003).

\bibitem{Luo04}
X. Luo, S.~B. Zhang, and S. H. Wei,
\newblock Phys. Rev. B {\bf 70}, 033308 (2004).

\bibitem{Silva04}
A.~J.~R. da Silva, A. Fazzio, and A. Antonelli,
\newblock Phys. Rev. B {\bf 70}, 193205 (2004).

\bibitem{Wu04}
H.~Wu, M.~Hortamani, P.~Kratzer, and M.~Scheffler,
\newblock Phys. Rev. Lett. {\bf 92}, 237202 (2004).

\bibitem{LaBella06}M.~R.~Krause, A.~J.~Stollenwerk, J. Reed,
V.~P.~LaBella, M. Hortamani, P. Kratzer, and M. Scheffler,
(submitted).

\bibitem{Perdew} J.P. Perdew, K. Burke, and M. Ernzerhof,
Phys. Rev. Lett. {\bf 77}, 3865 (1996).

\bibitem{Blaha} P. Blaha {\it et al.}, {\bf WIEN2k}, 2001. ISBN 3-9501031-1-2.

\bibitem{footnote} Structural relaxation breaks the local $T_d$ symmetry
and split the $t_2$ triplet of the Mn $3d$ orbitals into
$x^2-y^2$, $xz$ and $yz$, and the $e$ doublet into $3z^2-r^2$ and
$xy$ orbitals.

\bibitem{Note}
For isolated impurities in Si, the on-site Coulomb correlation
could be important. Our GGA+$U$ calculations for 1/4 ML Mn
$c(4\times 2)$ with Hubbard $U$=3~eV (Refs. \onlinecite{Picozzi04}
and \onlinecite{Ernst05}) and Hund exchange of 0.9~eV show that
the half-metallicity remains, whereas it is lost for $U$=5~eV, but
the spin-polarization is still higher than 40\%. This is also true
for the 1/4 ML-Mn $p(2\times 2)$ superstructure.

\bibitem{sasioglu:05}
E. {\c S}a{\c s}{\i}o{\u g}lu, I. Galanakis, L. M. Sandratskii,
and P. Bruno, J. Phys.: Condens. Matter {\bf 17}, 3915 (2005).

\bibitem{picozzi:06}
S. Picozzi, M. Le{\v z}ai\'c, and S. Bl{\"u}gel, phys. stat.sol.
(a) {\bf 203}, 2738 (2006).

\bibitem{Centoni05}
S. A. Centoni, B. Sadigh, G.~H. Gilmer, T.~J. Lenosky, T. D{\'i}az
de la Rubia, and C.~B. Musgrave, Phys. Rev. B {\bf 72}, 195206
(2005).

\bibitem{Ernst05}
A. Ernst, L.M. Sandratskii, M. Bouhassoune, J. Henk, and M. L\"{u}ders,
\newblock Phys. Rev. Lett. {\bf 95}, 237207 (2005).

\end{thebibliography}
\end{document}